\documentclass[10pt,conference]{IEEEtran}

\usepackage{cite}
\usepackage{amsmath,amssymb,amsfonts}
\usepackage{algorithm}
\usepackage{algpseudocode}
\usepackage{graphicx}
\usepackage{textcomp}
\usepackage{xcolor}

\usepackage{hyperref}
\hypersetup{
    colorlinks=true,
    allcolors=blue,
}
\newcommand{\email}[1]{\href{mailto:#1}{#1}}
\newcommand{\https}[1]{\href{https://#1}{#1}}

\usepackage{multirow, makecell}
\usepackage{hyphenat}

\usepackage{tikz}
\usetikzlibrary{arrows.meta, positioning}

\newcommand\latestDatasetLink{\https{doi.org/10.5281/zenodo.14245890}}

\begin{document}

\title{CARDS:
	A collection of package, revision, and miscellaneous dependency graphs}

\author{
	\IEEEauthorblockN{Euxane TRAN-GIRARD}
	\IEEEauthorblockA{\textit{LIGM, CNRS, Univ. Gustave Eiffel} \\
		Marne-la-Vallée, France \\
		\email{euxane.trangirard@univ-eiffel.fr} \\
		\https{orcid.org/0009-0003-4190-7151}}
	\and
	\IEEEauthorblockN{Laurent BULTEAU}
	\IEEEauthorblockA{\textit{LIGM, CNRS, Univ. Gustave Eiffel} \\
		Marne-la-Vallée, France \\
		\email{laurent.bulteau@univ-eiffel.fr} \\
		\https{orcid.org/0000-0003-1645-9345}}
	\and
	\IEEEauthorblockN{Pierre-Yves DAVID}
	\IEEEauthorblockA{\textit{Octobus S.c.o.p.} \\
		Montreuil, France \\
		\email{pierre-yves.david@octobus.fr} \\
		\https{orcid.org/0009-0005-3618-3560}}
}

\maketitle

\begin{abstract}
	CARDS
	(\emph{Corpus of Acyclic Repositories and Dependency Systems})
	is a collection of directed graphs which express dependency relations,
	extracted from diverse real-world sources such as package managers,
	version control systems, and event graphs.
	Each graph contains anywhere from thousands to hundreds of millions of
	nodes and edges,
	which are normalized into a simple, unified format.
	Both cyclic and acyclic variants are included (as some graphs, such as
	citation networks, are not entirely acyclic).
	The dataset is suitable for studying the structure of different kinds
	of dependencies, enabling the characterization and distinction of various
	dependency graph types.
	It has been utilized for developing and testing efficient algorithms
	which leverage the specificities of source version control graphs.
	The collection is publicly available at \latestDatasetLink.
\end{abstract}


\section{Introduction}

Software development involves dependency systems at several levels:
version-control repositories track dependency between successive versions of
the source code, and package managers track dependencies between libraries
in a given environment. Such systems maintain underlying (mostly acyclic)
directed graphs, steadily growing with each new version and every new package.
We present a Corpus of Acyclic Repositories and Dependency Systems (CARDS),
collecting both package and versioning graphs in a unified setting.

The corpus contains 11 graphs extracted from system package managers
(from Linux distributions such as Arch Linux, Debian, NixOS, etc.)
for a total of nearly 700 thousand nodes.
Similarly, 13 language-specific library dependency graphs are integrated
(from software library repositories like Maven, npmjs, etc.)
for a total of 6 million nodes.
The second category of graphs comprises revision graphs extracted from
several hundreds of thousands public repositories,
versioned under both the Git and Mercurial distributed version-control systems,
with diverse history sizes from thousands to millions of commits,
for a total of several hundred million nodes.
Finally, we also include 45 dependency graphs from altogether different
sources,
in order to extend the corpus with as many types of DAGs as possible:
messages and events in Matrix chat rooms,
scientific paper citations,
law references, etc.
See Table~\ref{tab:datasets_overview} for complete details.

Each graph is given in a simple plain text format, in topological order;
cycles are broken when necessary,
in which case raw cyclic graphs are also made available.

The CARDS collection is made available under the Open Database License
(ODbL~v1.0~\cite{license:odbl}).
The associated scripts and programs used to produce the dataset are made
available under the European Union Public Licence
(EUPL~v1.2~\cite{license:eupl}).
The latest version is available on Zenodo~\cite{zenodo} at \latestDatasetLink.

\subsection{Usage and Related Works}
Previous works extracting dependency graphs exist for the Maven Central
repository~\cite{amine_benelallam_2018_1489120},
patents and scientific paper
citations~\cite{source:tang2008arnetminer}\cite{source:snapnets},
as well as court case citations~\cite{source:caselaw_citation_graph}.
To the best of our knowledge, CARDS is the first corpus gathering such
dependency graphs from various sources in a unified setting,
along with topological orderings to simplify further research.

We expect this corpus can be used to develop specific algorithmic
solutions for package managers and version control systems.
In particular, a subset of the repositories presented here have already been
used in a benchmark for new reachability algorithms in DAGs,
specifically tailored for version-control systems~\cite{bulteau:hal-04778558}.

\subsection{Organization of the Manuscript}
Section~\ref{sec:datasets} presents each dataset included in the corpus
individually, with its processing specificities.
Section~\ref{sec:format} details the file format used for dependency files,
both raw (.deps) and topologically sorted (.tdag).
Section~\ref{sec:methods} give the methods used to process datasets in
general, and the topological sort algorithm in particular.
We conclude in Section~\ref{sec:conclusion} with possible future improvements
of the dataset.

\newcommand{\kindrow}[2]{\multirowcell{#1}{\parbox{1.5cm}{\nohyphens{#2}}}}
\begin{table*}[t]
	\caption{Overview of the CARDS datasets}
	\begin{center}
		\begin{tabular}{l|l|c|c|c|c|r|r|r}
			\textbf{Group}
			 & \textbf{Dataset}
			 & \textbf{Selection}
			 & \textbf{Snapshot}
			 & \textbf{Labels}
			 & \textbf{Cyclic}
			 & $|G|$
			 & $\overline{|V|}$
			 & $\overline{|E|}$
			\\
			\hline
			\kindrow{5}{System packages}
			 & Arch Linux (including AUR)
			 & Whole
			 & 2024-10
			 & Package name
			 & Yes
			 & 4
			 & 108k
			 & 516k
			\\
			 & Debian
			 & Whole
			 & 2024-10
			 & Package name
			 & Yes
			 & 1
			 & 63k
			 & 297k
			\\
			 & FreeBSD ports
			 & Whole
			 & 2024-10
			 & Package name
			 & No
			 & 2
			 & 36k
			 & 1M
			\\
			 & Homebrew formul\ae
			 & Whole
			 & 2024-10
			 & Package name
			 & No
			 & 3
			 & 7k
			 & 7k
			\\
			 & NixOS nixpkgs
			 & Whole
			 & 2024-10
			 & Hash-Name
			 & Loops
			 & 1
			 & 116k
			 & 509k
			\\
			\hline
			\kindrow{5}{Project dependencies}
			 & Maven Central
			 & Whole
			 & 2018-09
			 & Package name
			 & Yes
			 & 6
			 & 223k
			 & 21k
			\\
			 & npmjs
			 & Whole
			 & 2024-10
			 & Package name
			 & Yes
			 & 2
			 & 3M
			 & 19M
			\\
			 & Perl CPAN
			 & Whole
			 & 2024-11
			 & Package name
			 & Yes
			 & 1
			 & 35k
			 & 353k
			\\
			 & R CRAN
			 & Whole
			 & 2024-11
			 & Package name
			 & No
			 & 1
			 & 21k
			 & 8k
			\\
			 & Rust Crates
			 & Whole
			 & 2024-11
			 & Package name
			 & Yes
			 & 3
			 & 55k
			 & 326k
			\\
			\hline
			\kindrow{3}{Source versioning}
			 & Software Heritage Bitbucket Mercurial repositories
			 & Whole
			 & 2020-07
			 & Commit hash
			 & No
			 & 245k
			 & 1k
			 & 1k
			\\
			 & Software Heritage Bitbucket Mercurial wikis
			 & Whole
			 & 2020-07
			 & Commit hash
			 & No
			 & 81k
			 & 3
			 & 2
			\\
			 & Git large open-source repositories
			 & Subset
			 & 2024-10
			 & Commit hash
			 & No
			 & 123
			 & 166k
			 & 191k
			\\
			\hline
			\kindrow{1}{Events}
			 & Matrix large public rooms
			 & Subset
			 & 2024-11
			 & Event ID
			 & No
			 & 38
			 & 40k
			 & 40k
			\\
			\hline
			\kindrow{5}{Citations}
			 & AMiner citation
			 & Whole
			 & 2023-01
			 & In-dataset ID
			 & Yes
			 & 1
			 & 5M
			 & 37M
			\\
			 & Caselaw Access Project
			 & Whole
			 & 2024-11
			 & In-dataset ID
			 & Yes
			 & 1
			 & 7M
			 & 68M
			\\
			 & Légifrance LEGI
			 & Whole
			 & 2024-11
			 & In-dataset ID
			 & Yes
			 & 1
			 & 2M
			 & 20M
			\\
			 & OpenAlex works
			 & Whole
			 & 2024-11
			 & In-dataset ID
			 & Yes
			 & 1
			 & 263M
			 & 3B
			\\
			 & SNAP citations
			 & Subset
			 & multiple
			 & In-dataset ID
			 & Yes
			 & 3
			 & 1M
			 & 5M
			\\
		\end{tabular}

		The CARDS collection is divided into multiple Datasets corresponding
		to the origin of their graphs.
		The more general kind of origin is given in the Group column.
		The Selection columns indicates whether the dataset considers all or a
		selected subset of the origin.
		The Snapshot column corresponds to the date of data extraction from the
		origin.
		The Labels column indicates the type of the node identifiers.
		The Cyclic column indicates the presence of cycles in the source
		graphs.
		The number of distinct graphs is given for each group,
		along with their average node and edge counts before cycle elimination.
		These exclude additional combined graphs.

		\label{tab:datasets_overview}
	\end{center}
\end{table*}

\section{Datasets}
\label{sec:datasets}

\subsection{Arch Linux packages (including AUR)}
This dataset contains software package dependency graphs for the Arch Linux
distribution.
It includes packages from the core and extra Arch Linux base
repositories \cite{source:archlinux_packages},
as well as packages from the Arch User Repositories (AUR)
\cite{source:archlinux_aur}, combined due to the dependency of the latter over
the base repositories.
The package metadata from the base repository were retrieved from the Arch
Linux mirror as a tar archive of plain text package information,
converted to JSONL using a custom Python tar-stream parsing script.
The package list and metadata from the AUR were retrieved from a name index and
the Aurweb JSON-RPC API \cite{source:archlinux_aurweb_rpc} respectively,
using another custom Python script for generating batches of API queries.
The metadata from both sources were normalized into a common format using jaq
\cite{tool:jaq}.
The provider-dependent relations were then resolved using DuckDB
\cite{tool:duckdb} (taking into account package replacement).
Version constraints were not considered when resolving dependencies.
Distinct dependency graphs were derived for runtime, make, check, and optional
dependencies.
An additional complete dependency graph was also derived as their union.
Cyclic dependencies were eliminated when deriving the acyclic variants using
method~\ref{sec:toposort}.

\subsection{Debian packages}
This dataset contains the software package dependency graph for the Debian
Linux distribution.
The complete package index was obtained from a Debian CDN package mirror
\cite{source:debian_mirror} in its specific Package index file format
\cite{source:debian_format}.
The index was then parsed using a custom Python script to build a temporary
provider-dependent mapping,
from which the dependency graph was derived.
Cyclic dependencies were eliminated when deriving the acyclic variants using
method~\ref{sec:toposort}.

\subsection{FreeBSD ports}
This dataset contains software package dependency graphs for the FreeBSD
operating system.
The complete package index was obtained from the FreeBSD Ports and Packages
Collection FTP \cite{source:freebsd_ports_index} in its specific INDEX file
format \cite{source:freebsd_index_format},
from which dependencies were extracted using a custom Python parsing script,
then directly sorted in a topological order using the method~\ref{sec:toposort}
(without cycle removal).
Distinct dependency graphs were derived for runtime and build dependencies.
An additional graph was derived as the union of the two.

\subsection{Homebrew formul\ae}
This dataset contains software package dependency graphs of the Homebrew
package manager for macOS and Linux.
The complete package index was obtained from the Homebrew Formul\ae~
JSON API \cite{source:homebrew_formulae_api},
from which dependencies were extracted using jaq~\cite{tool:jaq}
and directly sorted in a topological order using the method~\ref{sec:toposort}
(without cycle removal).
Distinct dependency graphs were derived for runtime, build, and test
dependencies.
An additional graph was derived as the union of the three.

\subsection{NixOS nixpkgs}
This dataset contains the software package dependency graph from the nixpkgs
package collection.
The complete package specifications were obtained by cloning the nixpkgs git
repository \cite{source:nixpkgs_git},
from which attributes were enumerated and evaluated to obtain their respective
derivation output hash using Nix~\cite{DolstraJV04}.
This operation took approximately a week.
The referenced packages of each of those output packages were then queried from
the NixOS public cache to derive the dependency graph using jaq
\cite{tool:jaq}.
Loops were eliminated when deriving the topologically-sort graph
using~\ref{sec:toposort}.

\subsection{Maven Central}
This dataset contains Java and JVM-related software library dependency graphs
from the Maven Central repository.
The CSV dependency tables of all packages (called "artifacts" in the Java
ecosystem) from the Maven Central dependency graph dataset
\cite{DBLP:journals/corr/abs-1901-05392}\cite{amine_benelallam_2018_1489120}
were re-used.
The dependency relations were matched using DuckDB \cite{tool:duckdb},
producing one graph per scope of dependency (runtime, compile, system,
provided, test, and import).
Only the latest artifact version was considered for each package.
Artifact version constraints were not considered for resolving dependencies.
Cyclic dependencies were eliminated when deriving the acyclic variants using
method~\ref{sec:toposort}.

\subsection{npmjs}
This dataset contains the Node (JavaScript and related) software library
dependency graphs from the npmjs repository.
The full packages metadata were obtained by mirroring the full npmjs.com
registry from its CouchDB replication endpoint \cite{source:npm_registry_api}
using a custom Python script.
This operation took about a week, resulting in a 30GB (compressed) JSONL index.
Package dependencies were then extracted using jaq~\cite{tool:jaq}.
Only the latest version was considered for each package.
Version constraints were not considered for resolving dependencies.
Distinct dependency graphs were derived for runtime and development
dependencies.
An additional graph was derived as the union of the three.
Cyclic dependencies were eliminated when deriving the acyclic variants using
method~\ref{sec:toposort}.

\subsection{Perl CPAN}
This dataset contains the software library dependency graph from the Perl CPAN
repository.
The full metadata of the packages (called "distributions" in the Perl
ecosystem) were obtained from the MetaCPAN Elasticsearch API
\cite{source:metacpan_api_doc}
using a custom Python script, producing a JSONL index.
The lists of modules provided by each package were then extracted using
jaq~\cite{tool:jaq}.
Likewise for the module dependencies of each package.
The provided-dependent packages were then matched using
DuckDB~\cite{tool:duckdb},
ignoring version constraints.
Cyclic dependencies were eliminated when deriving the acyclic variants using
method~\ref{sec:toposort}.

\subsection{R CRAN}
This dataset contains the software library dependency graph from the R CRAN
repository.
The packages metadata were obtained by mirroring the package index on the
CRAN website \cite{source:cran_web} using wget~\cite{tool:wget},
and extracting the dependency information from the package pages using a
custom Python program to form a single graph.

\subsection{Rust Crates}
This dataset contains the software library dependency graph from the Rust
crates.io repository.
The packages metadata were obtained from the crates.io package index git
repository \cite{source:crates_index_git} as JSON manifests,
from which dependency relations were extracted using jaq~\cite{tool:jaq}.
Only the latest version was considered for each package.
Version constraints were not considered for resolving dependencies.
Distinct dependency graphs were derived for normal, development, and build
dependencies.
An additional graph was derived as the union of the three.
Cyclic dependencies were eliminated when deriving the acyclic variants using
method~\ref{sec:toposort}.

\subsection{Software Heritage Bitbucket Mercurial archive}
Bitbucket~\cite{source:bitbucket} is an online software forge which offered
hosting for projects using the
Mercurial~\cite{tool:mercurial}\cite{mackall2006towards}
version-control system.
The 254k Mercurial projects hosted on it were archived by Software Heritage
\cite{source:swh_bitbucket_archive}\cite{source:boatbucket} in 2020,
when support for Mercurial was dropped by the platform.
A local copy of this 4.7TB compressed snapshot was re-packed from .tar.gz into
one seekable .zip archive per repository.
81k projects were also accompanied by a wiki
(the vast majority of which only containing one revision),
also making use of Mercurial as version tracker.
The commit history was then obtained through the hg~\cite{tool:mercurial}
command line, reading the seekable archives through a virtual mount point
using mount-zip~\cite{tool:mount-zip}.
This IO-bound extraction took about a week.

\subsection{Git large open-source repositories}
This dataset contains the commit history graphs of a selection of 123 large
open-source software projects which use git~\cite{tool:git} for source
versioning.
The repositories were first cloned in full, resulting in a total of 170GB of
data.
The acyclic commit graphs (one per repository) were then extracted using the
git command line interface.

\subsection{Matrix rooms}
Matrix~\cite{source:matrix} is a decentralized communication protocol in which
events are stored in a directed acyclic graph.
Those graphs were obtained by joining 38 popular public chat rooms and
spaces (groups) on the Matrix network using a local instance of the Synapse
server~\cite{source:matrix_synapse}.
The graphs (one per room and one per space) were extracted from the PostgreSQL
database~\cite{source:matrix_synapse_dag_doc, tool:postgresql}.
While the graphs did not contain cyclic dependencies, the topological sort
needed to be recomputed~\ref{sec:toposort}, as the server storage ordering is
not necessarily topological (even taking into account the appropriate sort
keys).

\subsection{AMiner Citation}
AMiner is a published dataset of citations extracted from
DBLP~\cite{source:tang2008arnetminer}\cite{source:aminer_citation_v14}.
The citation network was extracted from the latest version of the dataset
(DBLP+Citation v14), given in 6GB compressed JSON data,
using jaq~\cite{tool:jaq}.
Cyclic dependencies were eliminated when deriving the acyclic variants using
method~\ref{sec:toposort}.

\subsection{Caselaw Access Project}
This dataset contains the citation network between the 6.9 million U.S. unique
court cases from the Caselaw Access Project (CAP)~\cite{source:caselaw}.
Because the pre-processed citation graph~
\cite{source:caselaw_citation_graph} was no longer available
for download at the time of writing, our citation graph was instead derived
from the source metadata instead.
The case metadata for each available volume was retrieved in bulk from the
CAP website~\cite{source:caselaw_doc} using wget~\cite{tool:wget},
totaling to 30GB of JSON data.
The citation network was then extracted from using jaq~\cite{tool:jaq}.
Cyclic dependencies were eliminated when deriving the acyclic variants using
method~\ref{sec:toposort}.

\subsection{Légifrance LEGI}
This dataset contains the full citation network between all the laws, codes,
and regulations of the French Republic available in consolidated format
(from 1945), from the Légifrance LEGI dataset \cite{source:legi}.
The LEGI law data were mirrored from the OpenData web server
\cite{source:legi_archive} as tar archives using wget \cite{tool:wget}.
The citation network was then extracted from the XML consolidated legal
documents using a custom Python script, skimming through the tar archives in
stream mode to filter the citations using an XPATH query \cite{tool:xpath}.
Cyclic dependencies were eliminated when deriving the acyclic variants using
method~\ref{sec:toposort}.

\subsection{OpenAlex works}
This dataset contains the full reference network between all the works in the
OpenAlex bibliographic
database~\cite{priem2022openalexfullyopenindexscholarly}.
The whole snapshot (about 390GB compressed) was retrieved in bulk from S3
as JSONL entries \cite{source:openalex_aws}.
The references were then extracted using jaq \cite{tool:jaq}.
This bulk retrieval and stream-processing was faster than using projections
through the OpenAlex API.
Cyclic dependencies were eliminated when deriving the acyclic variants using
method~\ref{sec:toposort}.

\subsection{SNAP citations}
This dataset contains three citations graphs from the SNAP
collection~\cite{source:snapnets},
for U.S. patents from 1963 to 1999~\cite{source:snap_patents}, and for
high-energy physics theory~\cite{source:snap_hepth}
and phenomenology~\cite{source:snap_hepph}
articles on arXiv~\cite{source:arxiv} between 1993 and 2003.
The SNAP graphs were retrieved in a format directly compatible with our own
.deps~\ref{sec:deps}.
Cyclic dependencies were eliminated when deriving the acyclic variants using
method~\ref{sec:toposort}.

\section{Data format}
\label{sec:format}

The graphs of the CARDS collection are encoded in the plain text formats
described in this section.

\subsection{Dependency specifications: .deps}
\label{sec:deps}
The .deps format is used to encode directed graphs.
Each line is a space-separated list of one or more identifiers,
specifying a node and nodes on which it depends.
Lines are ended with \verb|\n| and may appear in any order.
Identifiers can be any space- and line return-free strings.

Nodes and dependencies may be specified multiple times,
on the same line or over multiple lines.
The concatenation of multiple .deps files corresponds to the union of the
graphs described by each, merging dependencies.

\subsection{Topological DAG: .tdag}
\label{sec:tdag}
\begin{figure}
	\centering
	\begin{tikzpicture}[node distance=0.2cm, auto]
		\node (a) {a};
		\node (b) [below=of a] {b};
		\node (c) [below=of b] {c};

		\path[->] (b) edge node {} (a);
		\path[->] (c) edge node {} (b);
		\path[->] (c) edge [bend left=75] node {} (a);

		\node[right=1.5cm of a, align=left] (ta) {\texttt{a}};
		\node[right=1.5cm of b, align=left] (tb) {\texttt{b a}};
		\node[right=1.5cm of c, align=left] (tc) {\texttt{c b a}};
	\end{tikzpicture}

	\caption{
		An example DAG on the left and a corresponding .tdag on the right.
	}
	\label{fig:example_tdag}
\end{figure}
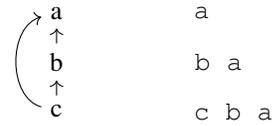

The .tdag format is a constrained variant of the .deps format for DAGs,
with exactly one line per node (no repetitions) and
nodes appearing in topological order.
See Fig~\ref{fig:example_tdag} for an example of DAG and its .tdag encoding.

The .tdag format allows consumer programs to incrementally load a graph's nodes
(with references only to known nodes at any given step).
The format is also incremental: new nodes depending only on existing ones can
be appended to the file.

In this collection, .tdag are derived from .deps using topological sort with
cycle elimination from~\ref{sec:toposort}.
The concatenation of multiple .tdag files results in a valid .deps file,
which can be canonicalized again using the same method.

\section{Processing methods}
\label{sec:methods}
\subsection{Topological sort with cycle elimination}\label{sec:toposort}

Topological sort was performed by a simple Depth-First Search.
Edges from the current node to another \emph{open} node
(i.e., with a known path to the current node) are removed,
thus effectively removing cycles.
This algorithm was chosen for its simplicity and linear time,
even though the result may not be a minimum-size feedback arc set.
This was implemented as a Nim program, provided together with the collection.

\subsection{General extraction and processing method}

The graphs were extracted using the following main steps, with
adaptations for each input as described in the previous section.
All scripts were run on local machines.

Data was acquired first using bulk data archives,
unless a fast API was available, using
curl~\cite{tool:curl} and wget~\cite{tool:wget}.
Flat files and archives were then processed as streams
(in various formats for intermediate files:
jsonl~\cite{format:jsonl}\cite{format:ndjson},
tsv, csv;
then deps~\ref{sec:deps} and tdag~\ref{sec:tdag})
then compressed using gzip~\cite{tool:gzip}.
For this processing, we used a combination of command line tools, general and
domain-specific languages:
jq with jaq~\cite{tool:jaq},
SQL with DuckDB~\cite{tool:duckdb},
Python~\cite{tool:python},
Nim~\cite{tool:nim},
as well as Fish shell~\cite{tool:fish} scripts to define the transformation
pipelines.

The main bottleneck was disk IO for most operations (excluding topological
sorting).
Disk seek operations and RAM usage were kept to a minimum by leveraging stream
processing,
including for tar and zip archives, allowing skipping unpacking their content.

All scripts are provided within the Zenodo archive:~\latestDatasetLink.

\section{Future works and improvements}
\label{sec:conclusion}
This dataset offers a large collection of acyclic graphs to study dependency
systems in software development.
A first possible improvement would be to run advanced feedback arc set
algorithms in order to maximize the number of retained dependencies in the
sorted DAGs, or maybe provide the graph of strongly connected components.
Also, other types of DAGs could be included,
in particular functional dependencies (function calls, type usage, ...)
in existing source code would be of interest.

\bibliographystyle{IEEEtran}
\bibliography{biblio}

\end{document}